\documentclass[9pt,twocolumn,twoside]{osajnl}

\journal{josab} 

\setboolean{shortarticle}{true}

\usepackage{relsize}
\usepackage{xcolor}
\usepackage{soul}
\usepackage[utf8]{inputenc}
\usepackage{textcomp}
\usepackage{newunicodechar}
\usepackage{amsmath} 
\usepackage{siunitx}
\usepackage{graphicx}
\usepackage{tabularx}
\usepackage[roman]{parnotes}

\title{High-performance, compact optical standard}

\author[1,$\dagger$,*]{Zachary L. Newman}
\author[1,$\dagger$]{Vincent Maurice}
\author[1,2]{Connor Fredrick}
\author[1]{Tara Fortier}
\author[1]{Holly Leopardi}
\author[3]{Leo Hollberg}
\author[1,2]{Scott A. Diddams}
\author[1]{John Kitching}
\author[1]{Matthew T. Hummon}

\affil[1]{Time and Frequency Division, National Institute of Standards and Technology, 325 Broadway, Boulder, CO 80305, USA}
\affil[2]{Department of Physics, University of Colorado, 2000 Colorado Ave., Boulder, CO 80309, USA}
\affil[3]{Department of Physics, Stanford University, 382 Via Pueblo Mall, Stanford, CA 94305-4060}
\affil[$\dagger$]{These authors contributed equally to this work.}
\affil[*]{Corresponding author: zachary.newman@nist.gov}

\begin{abstract}
We describe a high-performance, compact optical frequency standard based on a microfabricated Rb vapor cell and a low-noise, external cavity diode laser operating on the Rb two-photon transition at \SI{778}{nm}. The optical standard achieves an instability of \SI{1.8e-13}{}$\tau^{-1/2}$ for times less than \SI{100}{\second} and a flicker noise floor of \SI{1e-14} out to \SI{6000}{s}. At long integration times, the instability is limited by variations in optical probe power and the AC Stark shift.  The retrace was measured to \SI{5.7e-13} after 30 hours of dormancy. Such a simple, yet high-performance optical standard could be suitable as an accurate realization of the SI meter or, if coupled with an optical frequency comb, as a compact atomic clock comparable to a hydrogen maser.
\end{abstract}
\setboolean{displaycopyright}{false}

\begin{document}

\maketitle

Two-photon transitions in alkali and alkaline earth metals are considered promising candidates for simple, field-deployable optical frequency references and optical clocks \cite{Grynberg1977,Hall1989}. To first-order, Doppler-free spectroscopy can be accomplished by driving the transition using counter propagating beams at the same frequency, resulting in narrow lines. Because atoms in all velocity classes contribute to the signal and because two-photon transitions are typically observed in fluorescence through decay channels at a wavelength distinct from the driving field, measurements of such transitions can be performed with high signal-to-noise ratios. As a result, optical standards based on two-photon transitions can achieve excellent stability and accuracy.\par
The two-photon transition in Rb at \SI{778}{\nano\meter} was studied extensively as an optical frequency reference throughout the 1990s and early 2000s \cite{Nez1993a,Hilico1998b,Poulin2002}. The short-term frequency stability of these early tabletop systems \cite{Poulin2002,Hilico1998b} was measured at $\approx3\times 10^{-13}\tau^{-1/2}$ out to \SI{1000}{\second}. More recently, improved long-term performance has been demonstrated at $\approx4\times 10^{-13}\tau^{-1/2}$ out to \SI{10000}{\second} \cite{Martin2018}. Dichroic standards are also being developed, in which  much smaller detunings from the intermediate level result in larger signals, at the cost of increased light shift \cite{Perrella2019,Gerginov2018}. With the recent progress in miniaturization of frequency comb technology (both fiber and microresonator based) there has been renewed interest in the Rb two-photon transition for use as the local oscillator in a compact low power optical clock \cite{Newman2019,Martin2018}. To that end, our group recently demonstrated a miniature optical reference, with the optical bench having a total volume \SI{35}{\centi\meter^3}, and performance of $3\times 10^{-12}\tau^{-1/2}$, limited by the frequency noise of the semiconductor probe laser \cite{Maurice2020}.\par
In this Letter, we stabilize a near-infrared, external cavity diode laser (ECDL) to the two-photon transition in rubidium-85 confined in a microfabricated vapor cell. We achieve an instability of $1.8\times 10^{-13}\tau^{-1/2}$ from \SIrange[]{1}{100}{\second}, comparable to or outperforming previous measurements of the two-photon standard over these time scales.  In the current configuration, the laser achieves an instability of $\approx$\num{1e-14} at \SI{1000}{\second}.\par
\begin{figure}[htb]
\centering
\includegraphics[width=\columnwidth]{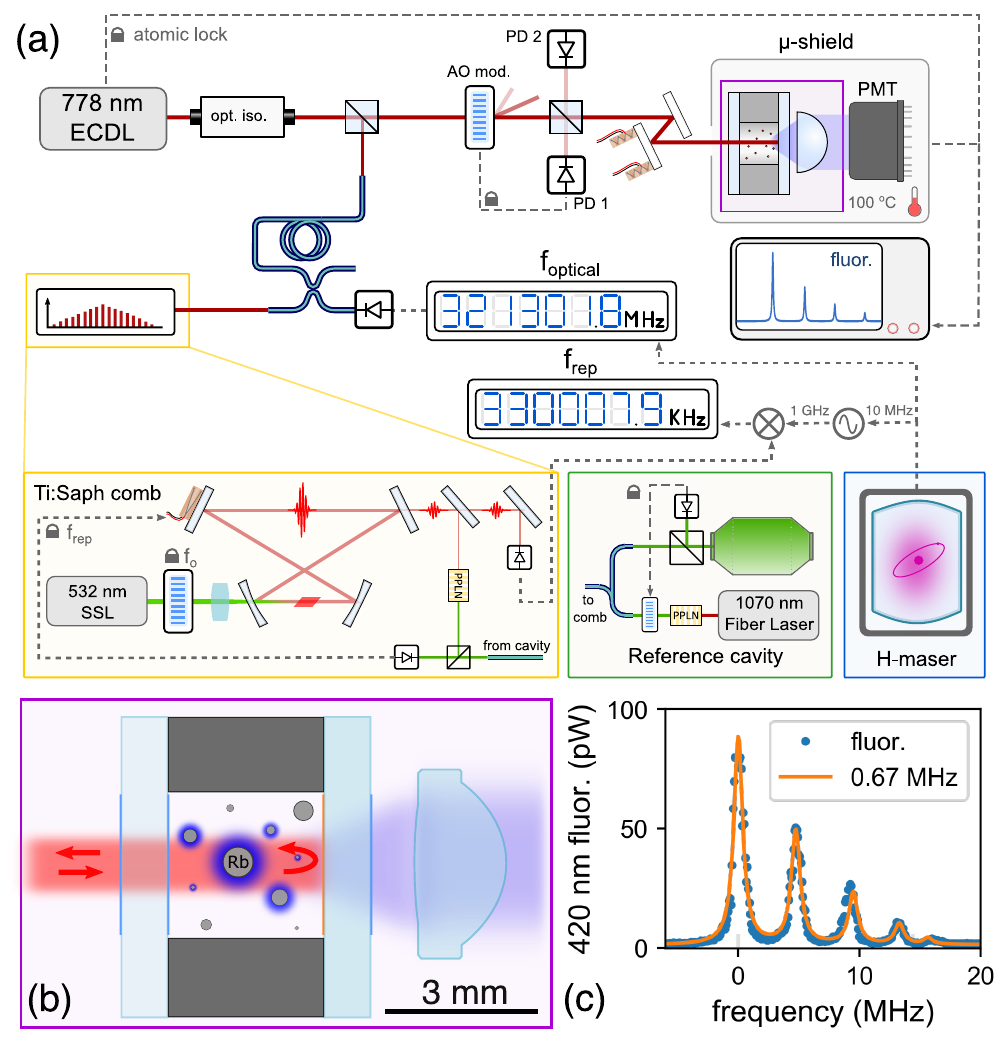}
\caption{\small
(a) The optical standard consists of a \SI{778}{\nano\meter} ECDL, power-stabilized using an acousto-optical modulator (AOM), a microfabricated vapor cell  housed in a magnetic shield and a photomultiplier tube (PMT). The laser is locked to the two-photon transition in Rb at 778 nm, and its frequency is measured with respect to a Ti:Sapph frequency comb stabilized to an optical cavity and hydrogen maser at short and long integration times, respectively. (b)  
The microfabricated vapor cell has optically coated windows and an aspheric collection optic directs the blue fluorescence from the atoms onto the PMT.  (c) 
Spectrum of two of the hyperfine components of the two-photon transition (blue) and the corresponding fit (orange) showing a linewidth of $\approx$\SI{670}{\kilo\hertz} FWHM.}
\label{fig:schematic}
\vspace{-4mm}
\end{figure}

Figure \ref{fig:schematic} shows a schematic of the optical standard which consists of an stabilized to the $5S_{1/2}\mathrm{(F=3)}\rightarrow5D_{5/2}\mathrm{(F=5)}$, two-photon transition in rubidium-85. Figure \ref{fig:schematic}(b) gives a close-up view of the vapor cell, which employs an in-line geometry for probing and measuring the two-photon transition \cite{Newman2019}, in which the counter-propagating beams necessary for avoiding Doppler-broadening of the transition are generated by retro-reflecting the laser off a high reflectivity dielectric coating on the back of a planar, microfabricated cell. We introduce rubidium into the cell by laser activatation of a rubidium dispenser compound after the cell is sealed. While this introduces some unwanted background gases, it serves as a convenient way to load atoms into a \SI{}{\milli\meter}-scale cell. The addition of a non-evaporable getter helps preserve good vacuum inside the chamber that allows us to do high-resolution spectroscopy and helps mitigate some of the shifts associated with background gases from the dispenser.\par
A photomultiplier tube placed at the back of the cell detects excitation of the transition via fluorescence at \SI{420}{nm}. In contrast, laboratory scale rubidium two-photon standards typically locate the PMT adjacent to the vapor cell \cite{Bernard2000,Poulin2002,Martin2018}. 
\color{black} The compact cell geometry also reduces the sensitivity of average probe beam power due to angular misalignment of the beam steering optics, as the sensitivity scales as roughly the square of the beam propagation length for the simple retro-reflection geometry \cite{Martin2019a}.\par
\color{black}
We measure the frequency of the ECDL by comparing the heterodyne beat between the ECDL and a mode of a self-referenced, \SI{1}{\giga\hertz} repetition-rate, titanium sapphire frequency comb \cite{Fortier2006} locked to an ultrastable optical cavity \cite{Young1999}. At short integration times the frequency comb stability, determined by the stability of the reference cavity ($\sigma \approx $\num{1e-15} at \SI{1}{\second}), is more than sufficient for measuring the frequency of the ECDL.  However, at long integration times linear drift of the cavity is a limiting factor in our measurements. To work around this issue, we track the cavity drift by measuring the repetition of the Ti:sapph laser against a maser-synthesized \SI{1}{\giga\hertz} tone and subtract a fourth-order polynomial fit of the frequency drift from the ECDL-comb beat in the optical domain.\par
Figure \ref{fig:schematic}(c) shows a spectrum (blue dots) of the hyperfine components ($F = 5$ to $1$) of the clock transition as the laser frequency is swept across the resonance with a fit (orange line) that gives a linewidth of \SI{670}{\kilo\hertz}. The observed linewidth includes contributions from the natural linewidth (\SI{330}{\kilo\hertz}), collisional broadening ($\approx$\SI{160}{\kilo\hertz}) from helium that diffuses into the cell through the glass windows, transit time broadening ($\approx$\SI{110}{\kilo\hertz}) resulting from the finite interaction time of the atoms with the laser \cite{Biraben1979}, the laser linewidth (\SI{39}{\kilo\hertz}) and $\approx$\SI{20}{\kilo\hertz} resulting from collisions with unwanted background gases that arise from introducing rubidium into the cell. These contributions are consistent with earlier measurements of similar microfabricated cells \cite{Newman2019}.\par
\begin{figure}[htbp]
\centering
\includegraphics[width=\columnwidth]{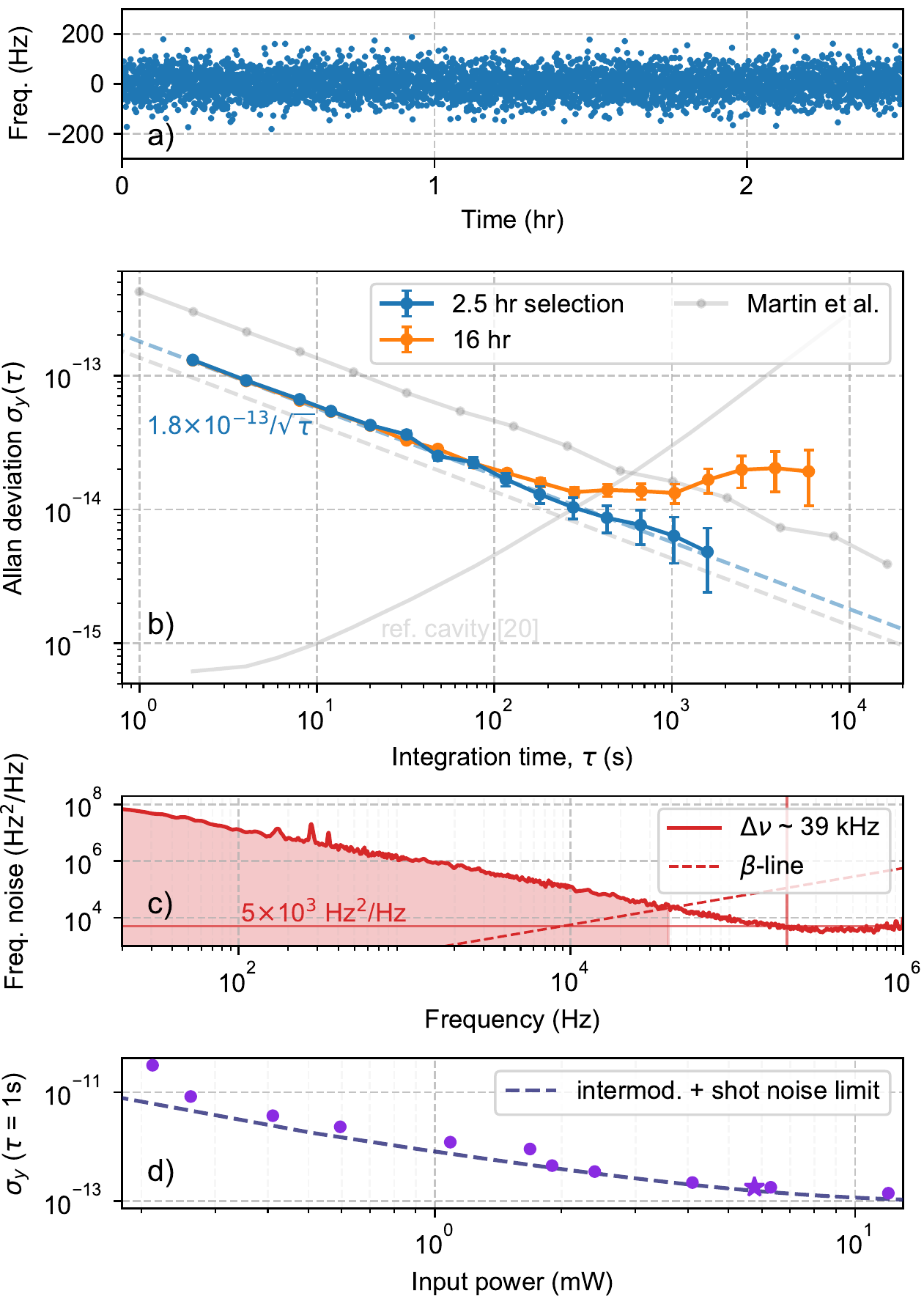}
\caption{\small Short-term frequency stability of the optical standard. (a) Time-series measurement of the beat frequency between the Ti:Sapph frequency comb and the ECDL. (b) Allan deviations of the rubidium-stabilized clock laser frequency with $1\sigma$ error bars for a \SI{16}{\hour} period (orange), the \SI{2.5}{\hour} selection (blue) shown in (a). (c) Frequency noise power spectrum of the free-running, unmodulated ECDL. The shaded line indicates the component of the spectrum that  contributes dominantly to the linewidth. The vertical red line indicates twice the modulation frequency. (d) Scaling of the \SI{1}{s} instability with input power.  Star denotes operating power for panels (a) and (b). The dashed line indicates a fit of the instability limit due to shot noise and intermodulation noise.}
\label{fig:short-term stability}
\vspace{-2mm}
\end{figure}

Figure \ref{fig:short-term stability}(a) shows a \SI{2.5}{\hour} time-series of the locked laser frequency from which we calculate the  Allan deviation of the optical standard, shown in Fig. \ref{fig:short-term stability}b. The typical performance of the system is well characterized by the \SI{16}{\hour} data set (orange). At longer times ($\approx$\SI{1000}{\second}) we observe a flicker noise floor of $\approx$\num{1e14}, and attribute this to ac Stark shifts caused by instability of the laser probe intensity as described in the following section. During time periods where the intensity of the driving field is more stable, however, we commonly see improved performance as exemplified by the \SI{2.5}{\hour} data set (blue).\par
We first consider the short-term instability, which scales as \SI{1.8e-13}{}$\tau^{-1/2}$ for times less than \SI{100}{\second}. There are two main contributions to this instability that we assess in panels (c) and (d).  Panel (c) shows the measured frequency noise spectrum $S_{\Delta\nu}(f)$,  of the free-running ECDL. According to \cite{Audoin1991}, the short term instability limit imposed by frequency noise of the local oscillator (intermodulation noise) is $\sigma_y \approx \frac{[S_{\Delta\nu}(2f_m)/f_0^2]^{1/2}}{2}\frac{1}{\sqrt\tau}$, where $f_m$ corresponds to the dither frequency used to generate the error signal via lock-in detection, and $f_0$ is the clock laser frequency. For our system, this corresponds to $S_{\Delta\nu} = $\SI{5e3}{\hertz^2/\hertz} and an instability of \SI{9e-14}{}$\tau^{-1/2}$.  For reference, integration of the free-running frequency noise of the laser below the $\beta$-line (shaded red region) yields a \SI{39}{\kilo\hertz} linewidth.\par
\vspace{-1mm}
\newpage
Figure \ref{fig:short-term stability}(d) shows the scaling of the short term instability at \SI{1}{\second} averaging time with input probe power $P_\textrm{in}$. For an instability limited by fluorescence photon shot noise, we expect the instability to scale as $P_\textrm{in}^{-1}$, which agrees well with the observations for powers less than \SI{10}{\milli\watt}. The dashed line in Fig \ref{fig:short-term stability}(d) is a fit to the instability data where a term that scales as $P_\textrm{in}^{-1}$ and a constant offset of \SI{9e-14}, corresponding to the intermodulation limit, have been added in quadrature.  This model fits the instability data well and indicates that operating at powers larger than \SI{10}{mW} does not further improve the short term instability due to the limit imposed by intermodulation noise for this laser.  The time series data presented in Fig. \ref{fig:short-term stability} are taken with an input power of $\approx$\SI{6}{\milli\watt}, as denoted by the star in panel (d).\par
\vspace{-1 mm}
\begin{figure}[htbp]
\centering
\includegraphics[width=\columnwidth]{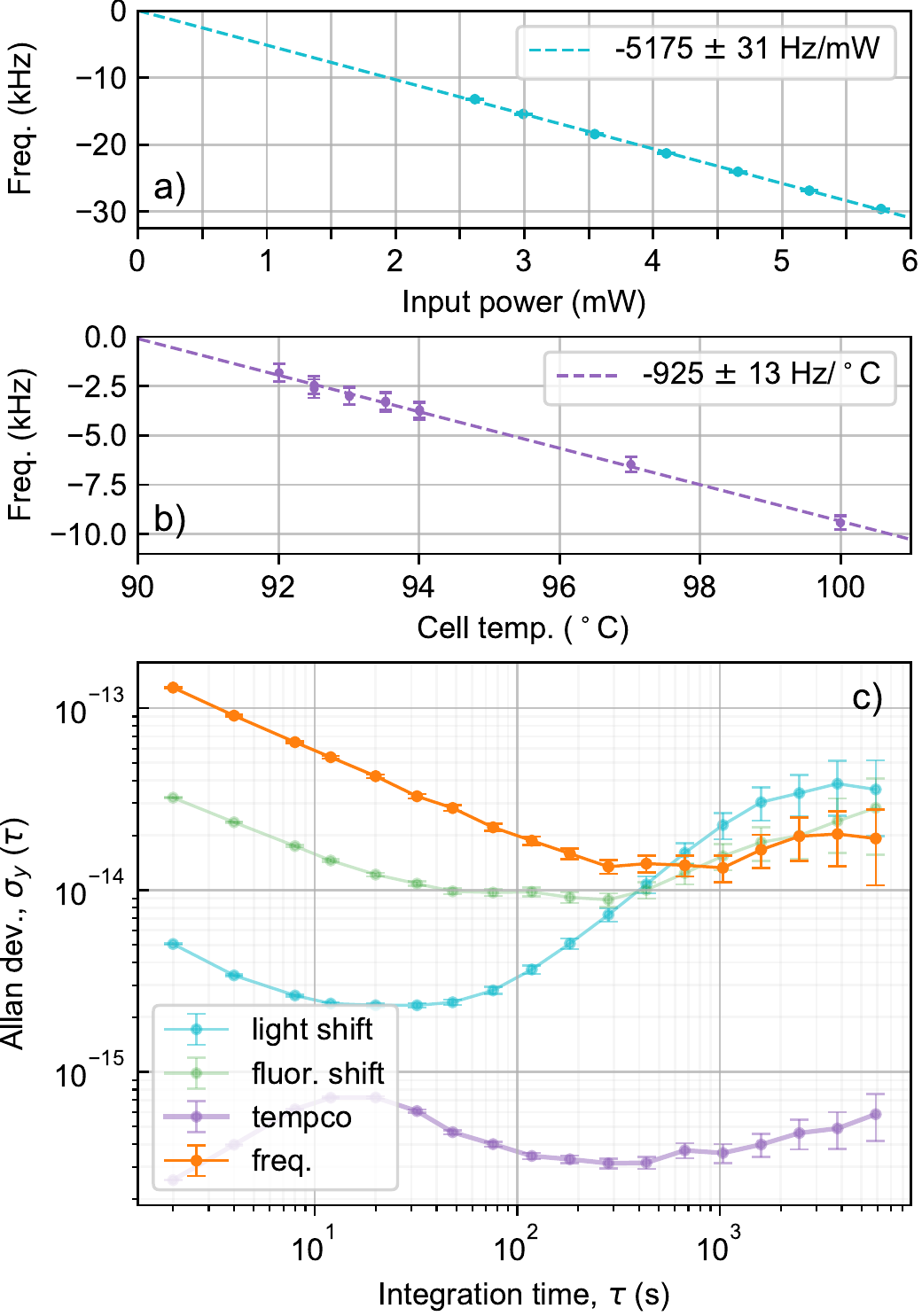}
\caption{\small Systematic contributions to the instability of the optical standard. (a) Frequency shift versus input power. (b) Frequency shift versus cell temperature. (c) Allan deviations showing the expected contribution to the clock frequency from the light shift (dark blue, green) and the collision shift (purple). $1\sigma$ error bars.}
\label{fig:systematics}
\vspace{-2mm}
\end{figure}
Figure \ref{fig:systematics} summarizes measurements of systematic effects that limit the stability of the optical reference at long times. Panels (a) and (b) show measurements of the ac Stark shift and vapor cell temperature shift, respectively. In panels (a) and (b), the input power and cell temperature are recorded using in-loop control  measurements.  Panel (c) shows the limits to the the laser instability imposed by the ac Stark shift and cell temperature variation as recorded by out-of-loop measurements for the laser power and cell temperature (PD2 as shown in \ref{fig:schematic}(b) and a auxiliary thermistor placed in the cell mount).\par
A measurement of the \SI{420}{\nano\meter} fluorescence amplitude, which is proportional to the square of the probe power, provides a second out-of-loop monitor of the laser power. Figure \ref{fig:systematics}(c), shows fractional instabilities for both these out-of-loop measurement techniques. We find that both of these out-of-loop power measurements overestimate the actual variation observed in operation of the optical standard at integration times longer than \SI{1000}{\second}.\par
\begin{figure}[htbp]
\centering
\includegraphics[width=\columnwidth]{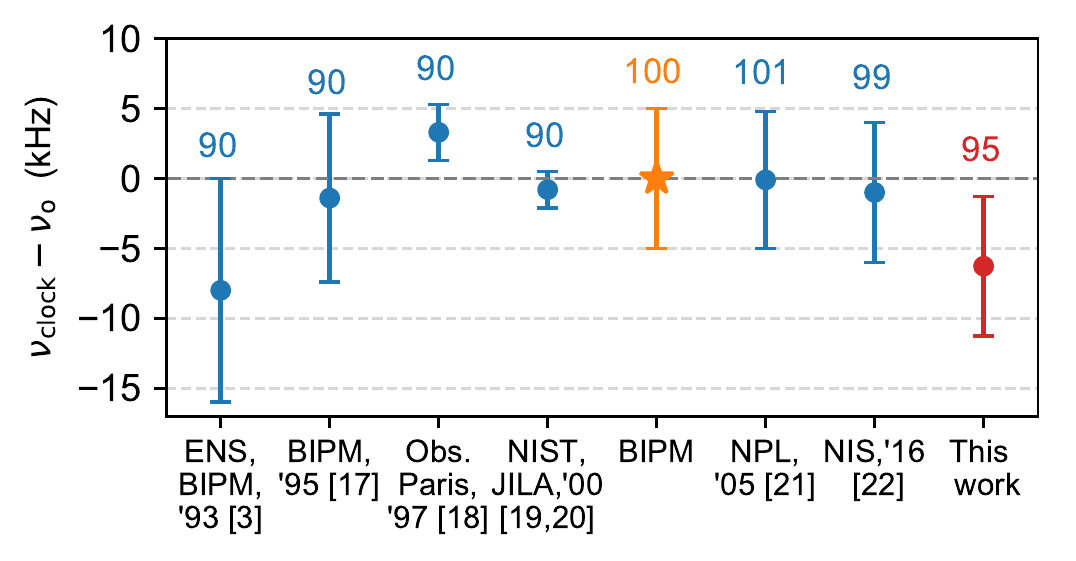}
\caption{History of measurements of the $^{85}$Rb two-photon transition. The orange star (dark gray dashed line) indicates the BIPM (2005) accepted value for the F=3$\rightarrow$F'=5 transition. The number above each measurement indicates the vapor cell temperature in \SI{}{\celsius}. Error bars for this work represent uncertainty due to operating temperature range. For previous measurements, refer to original papers for description of error bars.}
\label{fig:clock_freq}
\end{figure}

Figure \ref{fig:clock_freq} shows prior measurements of $^{85}$Rb, $F=3\rightarrow F'=5$ transition corrected for zero optical power \cite{Nez1993a,Felder1995,Touahri1997,Jones2000,Diddams2000,Edwards2005,Terra2016}. Taking the mean value of our 16 hour measurement campaign, and correcting for zero optical power and \SI{95} {\celsius} cell temperature, we measure the frequency of our optical standard to be $\approx$ (385 285 142 369 $\pm$ ~5)~\SI{}{\kilo\hertz} (red circle).  This corresponds to a frequency difference $\Delta\nu \approx$\ -6 $\pm$ 5 \SI{}{\kilo\hertz} from the BIPM (2005) accepted transition frequency \cite{Touahri1997,Jones2000}. The uncertainty of $\pm$~\SI{5}{\kilo\hertz} corresponds to the cell temperature shift over the typical operating temperature range of \SI{90} {\celsius} to \SI{100} {\celsius}. 

Frequency shifts due to collisions with background gas impurities typically limit the accuracy of vapor cell optical references \cite{Hilico1998b,Edwards2005}. Previous measurements made using fused silica or pyrex cells are likely to contain \SI{4}{mTorr} of helium due to permeation of atmospheric helium into the cell, leading to a frequency shift of $+4$~\si{kHz} compared to an evacuated cell \cite{Zameroski2014}.  In contrast, the vapor cell used in this work is fabricated using low helium permeability aluminosilicate glass windows, with a predicted helium leak rate \cite{Dellis2016} of \SI{1.2e-5}{Torr/year}, and negligible frequency shift due to collisions with helium.  Our rubidium cell loading method introduces additional background gases into the cell that we subsequently attempt to remove using pumping from a non-evaporable getter.  Thus, estimating the additional background gas impurities after the cell is sealed is difficult, but our two-photon linewidth and frequency measurements are consistent with pressures of less than \SI{2}{mTorr}.  
 \par

\vspace{-3mm}
\normalsize
\vspace{-2mm}

\begin{figure}[!t]
\centering
\includegraphics[width=\columnwidth]{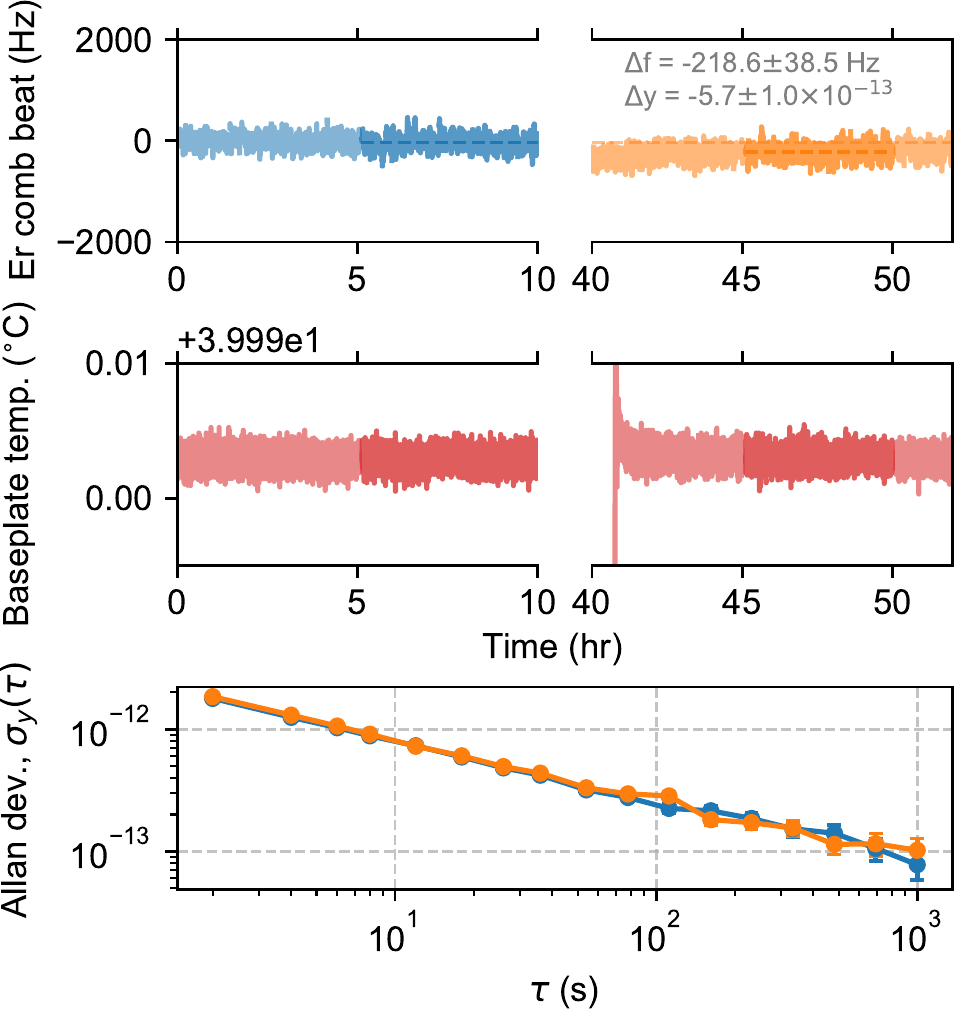}
\caption{\small Fractional retrace measurement showing the clock laser frequency (a) and cell housing temperature (b) over a $\approx$\SI{2}{\day} period. After \SI{10}{\hour} initial measurement, the laser and all control electronics are powered off, left in the "off" state for over a day and powered on again. (a) Beat note between the ECDL and the erbium fiber comb. Highlighted segments of the data indicate the portion of the data used to determine the retrace, allowing for a $\approx$\SI{4}{\hour} warm-up period over which environmental factors surrounding the cell equilibrate (the cell housing temperature is shown as an example). Dashed-dotted lines show the average frequency of the blue and orange measurement highlighted segments. (c) Fractional stability of the laser frequencies shown in (a). The fractional stability at \SI{1000}{\second}, $\approx$\SI{1e-13}{}$\tau^{-1/2}$ , corresponds to the uncertainty in our retrace measurements.}
\label{fig:retrace}
\vspace{-4mm}
\end{figure}

Figure \ref{fig:retrace} shows a \SI{24}{hr} retrace measurement. Retrace, as described by Vanier \textit{et. al.} \cite{Vanier2003}, in the context of our optical standard refers to a change in the clock laser frequency after the system power has been cycled and is particularly relevant for battery-powered or portable frequency standards. Figure \ref{fig:retrace} shows a fractional retrace measurement for the optical standard with a \num{5}-\num{24}-\num{5}-\num{5} \si{\hour} 'on'-'off'-'warmup'-'on' cycle.
After a warm up period of \SI{5}{h}, we achieve a retrace of $5.7\times10^{-13}$, which represents an improvement of close to $10^3$ compared to the retrace of the chip-scale atomic clock \cite{Microsemi}.  The warm up time of \SI{5}{h} is currently limited by thermalization of the optical bread board base plate, which could be reduced in the future by implementing a more compact physics package. This improvement in retrace will be important for applications where a portable clock needs to achieve a high level of accuracy after turn on when disciplining to a reference oscillator (e.g., GNSS timing) is impracticable.\par
In summary, we have demonstrated a high quality standard and presented a framework for future evaluations of compact optical references (absolute frequency combined with retrace measurements). Based on our measurements, we believe optical standards based on optical transitions in warm alkali vapors, such as rubidium, can support even higher performance; in fact we have observed \num{8e-14} at 1s using a low noise clock laser. We suspect short term instabilities at the \num{1e-14} level are within reach by utilizing high power, low noise laser in conjunction with a light shift mitigating interrogation scheme \cite{Yudin2020}.\par
\vspace{2mm}
\noindent\large{\textsf{\textbf{Funding information.}}}\normalsize 
~Defense Advanced Research Projects Agency (DARPA); Atomic Clocks with Enhanced Stability (ACES).
 \vspace{2mm}
 
\noindent\large{\textsf{\textbf{Acknowledgement.}}}\normalsize
~The authors would like to thank Samuel Brewer for his assistance in operating the optical reference cavity as well as Jenifer Black and Andrew Ferdinand for comments on the manuscript. The views, opinions and/or findings expressed are those of the authors and should not be interpreted as representing the official views or policies of the Department of Defense or the U.S. Government. Any mention of commercial products is for information only; it does not imply recommendation or endorsement by NIST. Distribution Statement "A" (Approved for Public Release, Distribution Unlimited).
\vspace{2mm}

\noindent\large{\textsf{\textbf{Disclosure.}}}\normalsize
~ZLN is a co-founder of Octave Photonics, a company specializing in nonlinear nanophotonics with applications for optical clocks.
\vspace{-3.5 mm}
\bibliography{references_old2}

\end{document}